\documentclass[
    ,final            
  ]
  {aipproc}
\layoutstyle{6x9}

\begin{document}

\title{Magnetogenesis from Rotating Cosmic String Loops}

\classification{98.80.Cq,98.54.Kt}
\keywords      {cosmic strings, loops, magnetogenesis}

\author{Thorsten Battefeld}{address={DAMTP, Center for Mathematical Sciences, University of Cambridge, Wilberforce Road, Cambridge,CB3 0WA, UK}}

\begin{abstract}
We present a mechanism to create vortices in a plasma via gravitational dragging behind rotating cosmic string loops. The vortical motions  create magnetic fields by means of the Harrison-Rees mechanism; the fields are further enhanced through galactic collapse and dynamo amplification. Employing the Velocity dependent One Scale model (VOS) for the string network and incorporating loop dynamics, we compute the magnetic fields generated around the time of decoupling: these are just strong and coherent enough to account for presently observed magnetic fields in spiral galaxies if efficient dynamos with $\Gamma_{dy}^{-1}\approx 0.3\,\mbox{Gyr}$ are present. 
\end{abstract}

\maketitle

\section{Introduction}
Magnetic fields in the $\mu\mbox{G}$ range have been observed in galaxies and clusters \cite{Widrow:2002ud}, though their origin remains a mystery. In the case of spiral galaxies, fields can be amplified considerably during the galaxy's numerous rotations  by means of a  dynamo; however, the difficulty of the problem still lies in the coherent and reasonably strong seed field requirement. 

Over the years, several scenarios generating seed fields have been proposed, but very few compelling explanations exist: astrophysical mechanisms operating within the horizon are not optimal for producing large coherence lengths.  Naturally, inflationary perturbations are present on the needed scales, but magnetogenesis is difficult to achieve due to conformal invariance in  electromagnetism \cite{Widrow:2002ud}.

In this note we summarize the proposal of \cite{Magnetogenesis}, explaining a possible mechanism to create plasma vortices via gravitational dragging in the vicinity of rotating cosmic string loops near the time of decoupling. These vortices subsequently cause magnetic fields through the Harrison-Rees mechanism \cite{Harrison,Rees}, which in turn are amplified during galaxy collapse and dynamo amplification. We try to be as conservative as possible regarding  the generation mechanism: we use string network parameters well within observational limits and do not assume turbulence, inverse cascades or large scale averaging. 

Previous work on magnetogenesis from string networks include \cite{Vollick:1993ac,Avelino:1995pm,Dimopoulos:1997df,Davis:2005ih}, which is extended in this study. We refer the reader to \cite{Widrow:2002ud,Grasso:2000wj,Semikoz:2005ks,Giovannini:2006kg} for general reviews on magnetogenesis, and to \cite{Hindmarsh:1994re,VilenkinShellard2000} for string networks. We set $c\equiv 1$ throughout.

\section{The Mechanism}

If a network of cosmic strings is present in the early universe, it has the intriguing feature to cause motions in the plasma on very large scales. This motion is caused by the gravitational interaction of strings with the plasma \cite{Vachaspati:1991tt,Vachaspati:1991sy,Vollick:1992sb}: if one averages over small scale wiggles on strings and considers only mildly relativistic velocities, a string acts as if it were a classical string with mass per unit length $\lambda$ in Newtonian gravity \cite{Vachaspati:1991tt,Vollick:1992sb} ($\lambda\approx 0.56 \mu_0$ in the matter era, where $\mu_0$ is the bare string mass density). Hence, a straight string moving with velocity $v_s$ causes a wake with in-falling velocity $v_i\sim G\lambda/v_s$ towards the sheet over-swept by the string, and dragging velocity $v_d\sim v_i^2/v_s$. The effect increases due to the contribution of the string's deficit angle for relativistic velocities \cite{Vachaspati:1991tt,Vachaspati:1991sy}, but we will be satisfied with the Newtonian estimate.

Vortices with $v_{rot}\sim v_i$ will arise on small scales due to turbulence in the wake, but it is rather questionable if these are large enough for magnetogenesis; on inter-string distances and without invoking turbulence, we expect vortices with $v_{rot}\sim v_d$ at most \footnote{It was assumed in \cite{Dimopoulos:1997df,Davis:2005ih} that vortices with $v_{rot}\sim v_i$ arise on inter-string distances, but no compelling argument was given for their presence -- see \cite{Magnetogenesis} for a more detailed analysis.}. 

However, stronger vortices arise naturally behind rotating loops due to gravitational dragging (still disregarding turbulence). Furthermore, loops are constantly produced in a network due to string intersection. Thus, the loop population will naturally lead to a wide variety of vortices, up to inter-string distances. As an illustration, consider a circular loop of length $\ell$, translational velocity $v_t$ and rotational velocity $v_r\perp v_t$; we can then estimate the magnitude of the angular velocity after the loop traversed to \cite{Magnetogenesis} 
\begin{eqnarray}
\omega_{pl}\sim \frac{v_d}{\ell}\sim\frac{v_i^2}{\ell v_r}\sim\frac{(2\pi)^2\lambda^2 G^2}{7\,\ell v_t^2v_r}\,.\label{omegapll}
\end{eqnarray}
This expression is valid for $v_t<v_r$, so that the plasma in the vicinity of the loop is exposed to a few rotations before the loop moves away. As expected, $\omega_{pl}$ is second order in $\lambda G$, since it is due to gravitational dragging. Although not considered here, the angular velocity of the plasma might be further amplified by turbulent effects.

Once a vortex is produced in the plasma, it can cause a magnetic field via the Harrison-Rees (HR) effect: Compton scattering on the CMB affects electrons and protons differently \cite{Rees}, leading to the build-up of a current and ultimately to a magnetic field of magnitude 
\cite{Avelino:1995pm}  
\begin{eqnarray}
B =\frac{2m}{e}\omega_{pl}\approx 10^{-4}\omega_{pl}\,,
\end{eqnarray}
where $B$ is in Gauss and $\omega_{pl}$ in $s^{-1}$. The HR effect ceases to be efficient around the time of decoupling $t_{dec}$.

As a result, vortices/magnetic fields produced during the matter era just before $t_{dec}$, which have the  largest possible size, are of prime interest to us.  After field generation, it redshifts as $B\propto a^{-2}$ due to flux conservation up until the protogalactic cloud collapses at around $z_{gf}\sim 10$. 

At this redshift, one needs at least $B\propto 10^{-29}\,\mbox{G}$ on $0.5-5\,\mbox{kpc}$ scales to account for magnetic fields in spiral Galaxies: during the collapse of the protogalactic cloud the field strength amplifies by a factor of order $8\times 10^{3}$ \cite{Widrow:2002ud}. If a very efficient dynamo amplification follows, for instance with a rate of  $\Gamma_{dy}^{-1}=0.3\,\mbox{Gyr}$ \cite{RST}, the field amplifies further by a factor of $\exp(\Gamma_{dy}^{-1}(t_0-t_{gf}))\approx 1.4\times 10^{19}$ up to today's micro-Gauss field strength. If dynamos are less efficient, as argued in \cite{Widrow:2002ud}, stronger seed fields are needed. 

\begin{figure}[t!] 
   \centering
   \includegraphics[width=0.5\textwidth]{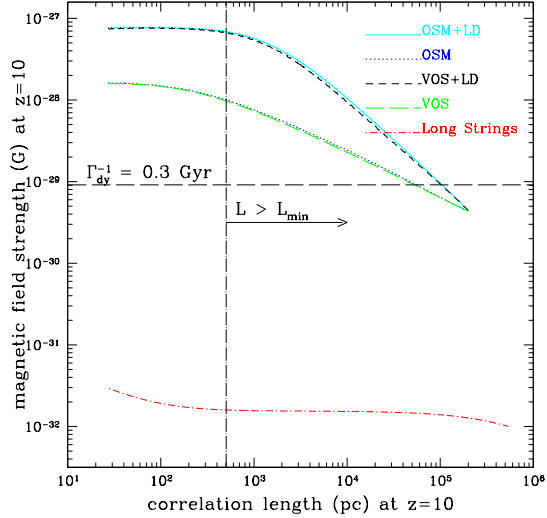} 
   \caption{Magnetic field strength at $z_{gf}=10$ for different string network models: OSM = One Scale Model, VOS = Velocity dependent One Scale model, with or without Loop Dynamics (LD); results are compared
   with the prediction from long string encounters, where the vortex is created in the region
   between two oppositely moving long strings. $L_{min}=0.5\,\mbox{kpc}$ covers only about $10\%$ of the protogalactic cloud and is the minimal coherence length needed. $\Gamma_{dy}^{-1}=0.3\,\mbox{Gyr}$ indicates the dynamo needed to amplify $B$ up to micro-Gauss field strength today -- dynamos below that line are strongly disfavored. In the numerical simulation we use  $G\mu_0=2\times 10^{-7}$, $\alpha=0.01$, $v_r(t_F)=0.4$  and $v_t(t_F)=0.1$.}
   \label{fig:bfield}
\end{figure}

How do magnetic fields behind loops compare to these needed seed fields? Simulating a cosmic string network by means of the velocity dependent one scale model \cite{Martins,Tye:2005fn}, as well as incorporating the loop population's evolution due to redshifting, emission of gravitational waves and dragging effects, leads to field strength of order $B\sim 5\times 10^{-28}\,\mbox{G}$ on kpc scales at a redshift of $z_{gf}=10$, see Figure \ref{fig:bfield}. Here, we use a conservative choice of network parameters \cite{Magnetogenesis}: $G\mu_0=2\times 10^{-7}$, initial loop length/horizon of $\alpha=0.01$, as well as initial loop velocities of $v_r=0.4$  and $v_t=0.1$. Even if we feed only $10\%$ of the energy transferred from the long string network into suitable large loops, fields cover already up to $10\%$ of the universe. Further, since loops accrete matter and are attracted to existing overdense regions as well, a $10\%$ coverage should be enough to cover most spiral galaxies. In addition, the coverage increases drastically if $v_t$ or $\alpha$ are increased mildly. For variations of parameters we refer the interested reader to \cite{Magnetogenesis}.

\section{Conclusion and Outlook}
In this note, we demonstrated a concrete mechanism to create vortices through gravitational dragging behind rotating string loops. Subsequently, magnetic fields are generated by the Harrison-Rees mechanism. The resulting fields are just strong and coherent enough to account for magnetic fields in spiral galaxies, if an efficient dynamo is present. However, magnetic fields in elliptical galaxies or clusters seem to be out of reach.

In case dynamos with rates $\Gamma^{-1}\approx 0.3\,\mbox{Gyr}$ are ruled out, or if $G\mu_0< 10^{-8}$ in future observations, our mechanism would be refuted. However, there is still the possibility that turbulent effects, which have been ignored entirely in this study, strengthen the existing vortices behind rotating loops. For instance, it is conceivable that turbulence redirects the in-falling plasma into the existing vortex, increasing the angular velocity and correspondingly the generated magnetic field.  

Thus, a numerical investigation into the flow patterns behind rotating loops is an interesting topic for future studies, especially if cosmic strings are observed in the near future.

\begin{theacknowledgments}
I would like to thank S.~Alexander for early discussions motivating this project as well as A.~C.~Davis and K.~Dimopoulos for feedback. Special thanks belong to my collaborators D.~Battefeld, M.~Wyman and D.~Wesley. T.~B. is supported by PPARC grant PP/D507366/1.\end{theacknowledgments}

\bibliographystyle{aipproc}   


\end{document}